\documentclass[twocolumn]{openjournal}
\usepackage{natbib}
\usepackage{graphicx,amsmath,amssymb,amstext}
\usepackage{amsbsy,amsfonts,amsthm,color}
\usepackage[colorlinks,linkcolor=blue,citecolor=blue,urlcolor=blue ]{hyperref}
\usepackage[utf8]{inputenc}
\usepackage{float}
\usepackage[caption=false]{subfig}
\newcommand{\cosech}{\mathrm{cosech} \,}
\providecommand{\del}[2]{\ensuremath{\frac{\partial #1}{\partial #2}}}

\begin{document}


\title{Evolution of Cosmic Voids in the Schr\"odinger-Poisson formalism}
\author{Aoibhinn Gallagher$^*$}
\email{$^*$Aoibhinn.Gallagher@mu.ie}
\author{Peter Coles}

\affiliation {Department of Theoretical Physics, Science Building, Maynooth University, Maynooth, Co. Kildare, Ireland.}


\begin{abstract}

We investigate the evolution of cosmic voids in the Schr\"odinger-Poisson formalism, finding wave-mechanical solutions for the dynamics in a standard cosmological background with appropriate boundary conditions. We compare the results in this model to those obtained using the Zel'dovich approximation. We discuss the advantages of studying voids in general and the advantages of Schr\"{o}dinger-Poisson description over other approaches. In particular emphasizing the utility of the free-particle approximation. We also discuss a dimensionless number, similar to the Reynolds number, for this system which allows our void solutions to be scaled to systems of different physical dimensions.
\end{abstract}

\maketitle

\section{Introduction}
\label{intro} 
Over the past few decades, galaxy redshift surveys, exemplified by those resulting from the Sloan Digital Sky Survey described by \cite{York_2000}, \cite{SDSS} and \cite{2017AJ....154...28B}, have revealed that the present-day Universe displays a rich hierarchical pattern of structure - the ``Cosmic Web'' - that encompasses a vast range of length scales, from individual galaxies to clusters and filaments surrounding enormous voids. On the other hand, observations of the cosmic microwave background, most recently from \cite{Planck_summary},  suggest that  the early Universe was almost homogeneous, with only slight temperature ﬂuctuations seen in the cosmic microwave background radiation. Models of structure formation link these observations through the eﬀect of gravity, relying on the fact that small initially over-dense regions accrete additional matter as the universe expands (a mechanism known as gravitational instability). The growth of density perturbations via gravitational instability is well understood in the linear regime, but the non-linear regime is much more complicated and generally not amenable to analytic solution. Numerical N-body simulations such as those discussed in \cite{Springel05} have led the way towards a fuller understanding of the structure formation process, but although such calculations have been priceless in establishing quantitative predictions of the large-scale structure expected to arise in a
particular cosmology, it remains important to develop as full an understanding as possible using analytical models. This is not only because simulation is not the same as comprehension but also because $N$--body methods are computational expensive which makes it difficult to explore a large parameter space using them alone. 

To complement the ``brute force'' approach of purely computational techniques, a variety of analytic techniques has been developed for describing various aspects of the evolution of density perturbations to various levels of approximation; for a good review of many of these approaches,see \cite{2002PhR...367....1B}. None of these capture all the details of the structure formation process, but each can provide useful information in particular circumstances. 

In this paper we study a particular formulation of the gravitational instability problem  - the Schr\"odinger-Poisson (SP) approach (explained below) -  which is based on wave mechanics rather than fluid mechanics.  This approach was first proposed in the cosmological context by \cite{widrowKaiser1993}, and has subsequently generated a great deal of interest; see e.g.  \cite{coles2002wave}, \cite{Szap}, \cite{Coles_2003}, \cite{CS2006a}, \cite{CS2006b} and \cite{johnston2009cosmological}. It turns out that this approach, while initially developed as an approximate method for handling the evolution of cold dark matter, is also applicable to a scenarios in which the behaviour of dark matter is inherently wave-like rather than particle-like, such as if the dark matter is a very light bosonic particle, perhaps in the form of a condensate, or some other form of ``fuzzy'' dark matter; see for example \cite{brook2009gravitational}, \cite{Schwabe_2020}, \cite{https://doi.org/10.48550/arxiv.2111.01199},  \cite{hui2021wave} and references therein.

The behaviour of cosmic voids provides an interesting test case for the SP system that complements the case of spherical collapse studied by \cite{johnston2009cosmological}. While gravitational collapse tends to amplify asymmetry, \cite{icke84} proved a ``Bubble Theorem'' according to which an initially asymmetric void becomes more spherical as the Universe evolves; see also numerical computations by \cite{Bert85}.  The study of a spherically symmetric void is therefore arguably more realistic than the collapse of a spherical over-density. Moreover, because matter tends to become compressed in sheets or filaments as the Cosmic Web develops, voids actually dominate space in terms of volume fraction. This is true for isolated voids, which is what we have done in this paper. As well as these physical aspects of void evolution there is an important statistical property shown by \cite{White79} that voids contain information about the entire hierarchy of $n$-point correlation functions at all orders. For these reasons and others there has been a great deal of recent interest in the behaviour of cosmic voids, see e.g.  \cite{Sheth_2004}, \cite{Bos_2012}, \cite{Achitouv_2015}, \cite{Demchenko_2016} and \cite{voids2019}.

Pulling together these chains of thought, in this paper we apply the Schr\"odinger-Poisson system to the case of isolated spherical voids. Although of course in a realistic context void expansion would be hindered by neighbouring structures, it is important to obtain as full an understanding as possible of the behaviour of the simplest case within the SP system. The layout of the paper is as follows: in Section \ref{cosmo}, we outline the background cosmology needed to build the results of this paper; in Section \ref{sp system}, we outline the basics of the SP system;  in Sections \ref{analytical_sols} and \ref{1dvoid}, we look directly at the void and apply what was covered in previous sections to solve for the dynamics of the void. In Section \ref{viscosity} we discuss scaling properties of the solutions in terms of an effective Reynolds number.  In Section \ref{conclusion} we present our conclusions and suggest future applications of our results and the approach that leads to them.

\section{Background Cosmology}
\label{cosmo} 

The current standard model of cosmology assumes the Cosmological Principle, according to which the universe is assumed to be homogeneous and isotropic, at least on large scales. Space-times consistent with this are described by the Robertson-Walker metric: 
\begin{equation}
    d s^2 = c^2 d t^2 - a^2(t)\left( \frac{d r^2}{1-\kappa r^2} + r^2d \theta^2 +r^2 \sin^2(\theta) d \phi^2 \right),\label{eq:RWmetric}
\end{equation}
where $\kappa$ is the spatial curvature: $\kappa=0$ represents flat space; $\kappa = +1$ represents constant positively curved space (closed universe); and $\kappa = -1$ represents constant negatively curved space (open universe). The time coordinate $t$ is cosmological proper time and $a(t)$ is the cosmic scale factor. The time evolution of the cosmic scale factor, $a(t)$, is determined via Einstein's gravitational field equations through the Friedman equation,
\begin{equation}
    3\left( \frac{\dot{a}}{a} \right)^2 = 8\pi G \rho - \frac{3\kappa c^2}{a^2} + \Lambda \label{eq:Freidman}
\end{equation}
the deceleration equation,
\begin{equation}
    \frac{\ddot{a}}{a} = -\frac{4 \pi G}{3}\left( \rho + 3\frac{p}{c^2} \right) + \frac{\Lambda}{3} \label{eq:deceleration}
\end{equation}
and the density-pressure relation,
\begin{equation}
    \dot{\rho} = -3\frac{\dot{a}}{a}\left( \rho + \frac{p}{c^2} \right) \label{eq:einstein3}
\end{equation}
where the dots denote derivatives with respect to cosmological proper time $t$, thus describing the global expansion or contraction of the universe. These models can be further parameterised by the Hubble parameter $H =\dot{a}/a$ and the density parameter $\Omega = 8\pi G \rho/3 H^2$. As usual the present epoch is defined by $t = t_0$, when $H = H_0$ and $\Omega = \Omega_0$. 

Throughout this paper we assume Newtonian gravity, since the scale of the perturbations are much smaller than the effective cosmological horizon $d_H = c/H$. We also do all calculations in co-moving coordinates, 
\begin{equation}
    \textbf{x} \equiv \textbf{r}/ a(t), 
    \label{eq:comoving}
\end{equation}
where $\textbf{x}$ is the co-moving spatial coordinate ($\textbf{x}$ is fixed in the frame of the Hubble expansion), $\textbf{r}$ is the spatial coordinate in our reference frame, and $a(t)$ is the cosmic scale factor.

\subsection{The Zel'dovich Approximation}
\label{zeldy} 

A simple and elegant approximation to the non-linear stage of gravitational evolution is the Zel'dovich approximation, introduced by \cite{zeldovich}. The initial density distribution is considered homogeneous and collisionless. The initial (Lagrangian) coordinates are $\textbf{q}$, then the Eulerian coordinates at time $\textbf{t}$ are given by 
\begin{equation}
    \textbf{r}(\textbf{q}, t) = a(t)[\textbf{q} + b(t)\textbf{s}(\textbf{q})], 
    \label{eq:zeldovich_approx}
\end{equation}
where $a(t)$ is the cosmic scale factor, $b(t)$ describes the evolution of a perturbation in the linear regime, and is defined by 
\begin{equation}
    \ddot{b} + 2 \Big(\frac{\dot{a}}{a}\Big) \dot{b} - 4\pi G \rho b = 0.
    \label{eq:btime}
\end{equation}
In a flat matter-dominated universe $b \propto t^{2/3}$. $\textbf{s}(\textbf{q})$ here is the displacement term, described by the initial velocity potential $\phi_0$ by, 
\begin{equation}
    \nabla \phi_0(\textbf{q}) = \textbf{s}(\textbf{q}).
    \label{eq:zel_velocity_potential}
\end{equation}
Mass conservation requires $\rho (\textbf{r}, t) d\textbf{r} = \rho_0 d \textbf{q}$, so the density field as a function of Lagrangian coordinates is
\begin{equation}
    \rho(\textbf{q}, t) = \rho_0 \left| \del{\textbf{q}}{\textbf{r}} \right| = \overline{\rho} \left| \delta_{ij} - b(t) \del{s_i}{q_j}  \right|^{-1}
    \label{eq:zel_rho_q}
\end{equation}
where $\overline{\rho} = (a_0/ a)^{3} \rho_0$. Here $\del{\textbf{r}}{\textbf{q}}$ is the deformation tensor, which accounts for the gravitational evolution of the density field. Due to the nature of $\textbf{s}(\textbf{q})$, the deformation tensor is a real symmetric matrix with eigenvectors that define a set of three principal (orthogonal) axes. Rewriting Equation (\ref{eq:zel_rho_q}) in terms of the eigenvalues $-\alpha(\textbf{q})$, $-\beta(\textbf{q})$ and $-\gamma(\textbf{q})$, we get 
\begin{equation}
    \rho(\textbf{q}, t) = \frac{\overline{\rho}}{[1- b(t)\alpha(\textbf{q})][1 - b(t)\beta(\textbf{q})][1 - b(t)\gamma(\textbf{q})]}
    \label{eq:zel_rho_2}
\end{equation}
If we use comoving coordinates (Equation (\ref{eq:comoving})) in 1 dimension, we get the much simpler expression for $\rho$ 
\begin{equation}
    \rho(q, t) = \frac{\rho_0}{1-b(t)\del{s(q)}{q}}
    \label{eq:zel_rho_final}
\end{equation}

The Zel'dovich approximation is an excellent tool for studying structure formation and is highly accurate up to the point of shell crossing \cite{osti_6192112}, at which point it breaks down. To combat this we need a model that goes beyond shell crossing, this is one reason why we look to the SP model. Other methods for studying multi-stream flow can be seen in \cite{multistream} and \cite{Gough2022}. 

\section{Schr\"odinger Poisson Regime}
\label{sp system} 

We start with the Newtonian equations for a self-gravitating perfect fluid.
\\
The \textit{Euler Equation}
\begin{equation}
    \del{\textbf{v}}{t} + (\textbf{v}\cdot \nabla ) \textbf{v} + \frac{1}{\rho} \nabla p + \nabla V = 0 \label{eq:euler},
\end{equation}
the \textit{Continuity Equation}
\begin{equation}
    \del{\rho}{t} + \nabla \cdot(\rho \textbf{v}) = 0 \label{eq:continuity},
\end{equation}
and the \textit{Poisson Equation}
\begin{equation}
    \nabla^2 V = 4\pi G \rho
    \label{eq:poisson}.
\end{equation}
Here \textbf{v}, $p$ and $\rho$ are the (peculiar) fluid velocity, pressure and density, respectively, and $V$ here is the gravitational potential. We simplify our treatment by assuming a cold (pressureless), irrotational fluid moving under the influence of a general potential {$V$}. This means that Equation (\ref{eq:euler}) now becomes the much simpler \textit{Bernoulli equation}

\begin{equation}
    \del{ \phi}{t} + \frac{1}{2}(\nabla \phi)^2 = -V \label{eq:bernoulli}, 
\end{equation}
where $ \textbf{v} = \nabla \phi, $ where $\phi$ is the velocity potential. 

We now make the \textit{Madelung Transformation}
\\
$\psi = \alpha e^{i\phi/\nu} $, where $\rho = \psi \psi^* = \alpha^2$. This gives 
\begin{equation}
    i \nu \del{\psi}{t} = -\frac{\nu^2}{2} \nabla^2 \psi + V \psi + P \psi \label{eq:schrodgen}
\end{equation}
where 
\begin{equation}
    P = \frac{\nu^2}{2} \frac{\nabla^2 \alpha}{\alpha}, \label{eq:QMpressure}
\end{equation}
from Equations (\ref{eq:bernoulli}) and (\ref{eq:continuity}). 
Equation (\ref{eq:schrodgen}) is  Schr\"odinger Equation, with potential $V$, $\nu$ acting as $\hbar$ with the addition of non-linear term $P$. Although $\psi$ is governed by the same equation as the evolution of a single-particle wave function, that is not how it should be interpreted. In particular, $|\psi|^2$ represents a physical density not a probability density, and its evolution is completely unitary - there is nothing like the wave-function collapse that occurs in standard quantum mechanics in this system.

It's important to note a crucial advantage of this description, namely that because $\rho=|\psi|^2$ the condition that $\rho\geq 0$ is automatically enforced if one applies, e.g., perturbation theory to $\psi$. This is not the case for approaches based on standard Eulerian perturbation theory applied to $\delta=(\rho-\rho_0)/\rho_0$ which predict $\rho<0$ when $\delta<-1$ and are therefore very unsuitable for describing voids. Note also that because the wavefunction describes a delocalized particle there are no singularities analogous to the caustics that form in the Zel'dovich approximation.

In the cosmological context we take $V$ to be the gravitational potential determined via the Poisson Equation and we get a system of coupled Partial Differential Equations (PDEs). We set $\nu$ to be an adjustable parameter so we have some control over the frequency of any oscillatory solutions that could arise from this system. $\nu$ has dimensions such that $\phi / \nu$ is dimensionless, $\phi$ is a velocity potential so has dimensions $L^{2} T^{-1}$, these are also the dimensions of kinematic viscosity, so $\nu$ can be thought of as a viscosity parameter for the model; we return to this later.

The original intentions of this model were to find a fluid interpretation of a quantum system, \cite{widrowKaiser1993} first proposed this be used to simulate cold dark matter (CDM). \cite{widrowKaiser1993} start with just the gravitational potential and not yet the fluid approach. It has been used predominantly in a cosmological context since. The quantum nature of this model is also an advantage as it gives us a new perspective on the behaviour of large scale structure. This is especially useful when one seeks to look beyond shell crossing, as this model can handle multiple streams.

\section{Cosmological Solutions}
\label{analytical_sols} 

Analytic solutions to the SP system can be obtained straightforwardly for the simplest case of a homogeneous and isotropic fluid. For this $P \equiv 0$, since $|\psi|^2 = \alpha^2 = \rho$ is defined to be homogeneous. We can start our study of a void by using solutions for an open and flat universe for the interior and exterior of the void in order to make a snapshot of a simple void configuration as follows.

\subsection{Spatially flat universe}
\label{flat} 

For the following we assume a flat universe (with $\kappa = 0$). We begin here by taking the well known cosmological result for the evolution of the density field under a globally defined cosmic time, $t$, given by

\begin{equation}
\rho(t) \equiv |\psi|^2 = \frac{\Lambda c^2}{8 \pi G} \text{cosech} ^2 \Big(\frac{3}{2}\sqrt{\frac{\Lambda c^2}{3}} t\Big).
\label{eq:rho_flat}
\end{equation}
Under spherical symmetry the Laplacian operator is simply
\begin{equation*}
\nabla^2 \equiv \del{^2}{r^2} + \frac{2}{r}\del{}{r},
\label{laplacian_r}
\end{equation*}
where r here is now the radial variable $r = |x|$. 
This density evolution obtained previously can be substituted into Equation (\ref{eq:poisson}) to obtain an equation for the gravitational potential $V$. This allows the calculation of $\phi$ by substitution into Equation (\ref{eq:schrodgen}), resulting in the wave-function $\psi$, as shown in \cite{johnston2009cosmological}:
\begin{equation}
V = \frac{\Lambda c^2}{12} (\text{coth}^2(\lambda t) -3)r^2
\end{equation}
\begin{equation}
\psi = \sqrt{\frac{\Lambda c^2}{8 \pi G}} \cosech (\lambda t) \exp \Big(\frac{i}{v} \sqrt{\frac{\Lambda c^2}{12}} \coth \left[\lambda t\right]r^2\Big)
\end{equation}
where $\lambda  \equiv \frac{3}{2} \sqrt{\Lambda c^2 / 3}$. We therefore also find
\begin{equation}
\phi = \sqrt{\frac{\Lambda c^2}{12}} \coth (\lambda t)r^2.
\end{equation}
Also following \cite{johnston2009cosmological}, a series expansion of $\rho = \alpha^2$ and $\phi/r^2$ about $\Lambda = 0$ provides the well known results of $\rho \propto t^2$ and $V = r^2/9t^2$ for this particular case.

\subsection{Open universe}
\label{open} 

Working in a similar way to Section \ref{flat}, we find solutions for a universe with negative curvature, $\kappa = -1$, an open universe, to be
\begin{equation}
  \rho =   \frac{ \gamma}{A^3(\cosh(\eta) - 1)^3},
\end{equation}
\begin{equation}
    V =  \left\{\frac{ G \pi \gamma}{12A^3\left[\sinh\left(\frac{\eta}{2}\right)\right]^3}
    \right\} r^2,
\end{equation}

\begin{equation}
\psi  = \frac{C}{\sqrt{2A}\left[\sinh\left(\frac{\eta}{2}\right)\right]^3} \exp \left[\frac{i}{v} C \coth \left(\frac{\eta}{2}\right) \cosech \left(\frac{\eta}{2}\right) r^2\right],
\end{equation}
where $A \equiv 4 \pi G \rho_0 /3$, $C = \gamma^{1/2}/4A$, $\mu = 3M/4\pi$ and $\gamma$ is a constant. For a simpler derivation we have converted to conformal time by using $ ad\eta = dt$ where $a$ is the scale factor of the open universe. 

Having established this simple solution we can now proceed to study the dynamics of a void, taking into account the fact that the density surrounding the void will not remain homogeneous as the interior expands.

\section{Modelling an Underdensity}
\label{1dvoid} 

In this Section we look at simple cases of a void growing under the gravity from its boundary. We will use the solutions found in Section \ref{analytical_sols} to form an analytic model of a void, in analogous fashion to the spherical collapse model studied by \cite{johnston2009cosmological}. We will also discuss the results of the numerical solutions to the SP system of equations for a simple one dimensional void model, both `free-particle' and including the gravitational potential in the evolution. These solutions will be compared to the Zel'dovich approximation for the same initial conditions, which is exact up to the moment of shell crossing, \cite{osti_6192112}. We see that shell crossing happens rather quickly in the case of a void though the formation of multi-stream regions forming at the edge of the void. This means that the Zel'dovich approximation applies for a shorter portion of the evolution in the void case than in the collapse case. 

\subsection{Analytic solutions to the void problem}
\label{void_analytic} 

\begin{figure}
    \centering
    \includegraphics*[width=.7\linewidth]{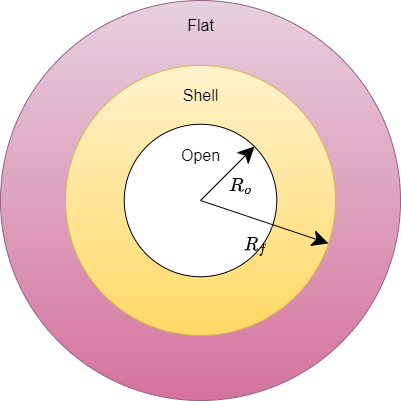}
    \caption{An illustration of the void, shell and flat space in white, yellow and pink respectively. It shows $R_{o}$ as the radius to the inner shell radius and $R_{f}$ as the radius of the outer shell.}
    \label{fig:void_sphere}
\end{figure}

Using the results from Section \ref{analytical_sols} and following the same method as described by \cite{johnston2009cosmological}, we can construct a simple model of a void. We construct the model of the void with an inner region and an outer fluid separated by a shell. The inner fluid is modeled by the open universe solutions and the outer fluid is modeled by the flat universe solutions. Both fluids are homogeneous but the fluid in the inner region is less dense than the outer fluid so expands more quickly, pushing into the surrounding medium and evacuating the void. This creates a small region on the cusp where the density is higher than the outer region. An illustration of how this would look is shown in Figure \ref{fig:void_sphere}. This figure shows the inside of the void in white, this is the area that gets emptied out. The yellow area is the shell of the void, this is where the denser region is created at the edge of the void. The pink area is the the rest of the universe which extends to infinity.

\begin{figure}
    \centering
    \includegraphics[width=0.9\linewidth]{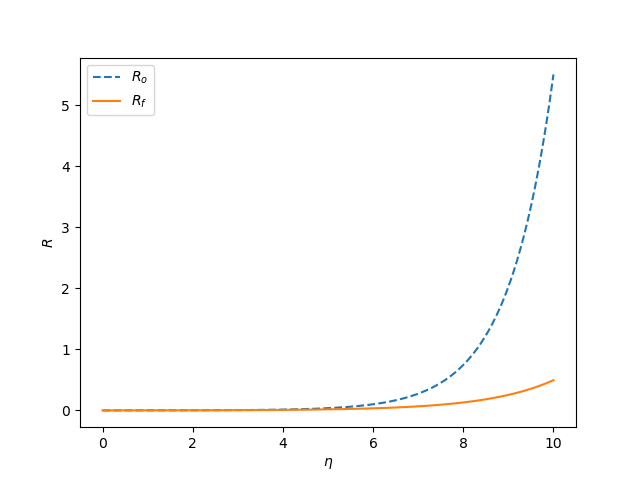}
    \caption{Evolution of the radii of both the inner section and outer section of the void, with respect to conformal time $\eta$.}
    \label{fig:radii_time}
\end{figure}

Using the cosmological solutions shown in Section \ref{analytical_sols}, and following the same procedure as \cite{johnston2009cosmological}, we compute an expression for the expansion of the radii of both sections of the void, in conformal time $\eta$. We have $R_{o}$ for the open-universe section of the void and $R_{f}$ for flat universe section of the void. 

\begin{equation}
R_o = \mu^{1/3} \frac{2A}{\gamma ^{1/3}}(\sinh^2(\eta/2)),
\end{equation}
and
\begin{equation}
R_f = B\left(\sinh(\eta) - \eta\right)^{2/3},
\end{equation}
where $B$ is a free parameter corresponding to the initial conformal time $\eta_0$ when the configuration is set up, at which point $R_o=R_f$. The evolution is illustrated by Figure \ref{fig:radii_time} which shows the evolution of each of these radii with respect to conformal time with an arbitrary scaling on the vertical axis just to show that the inner region of the void expands much more quickly than the outer region, and an overlapping region develops after some time as the inner void expands into a region that was previously in the space exterior to the void.

\subsection{The `free particle' void}
\label{void_numerical} 

\begin{figure}
    \centering
    \includegraphics*[width=\linewidth]{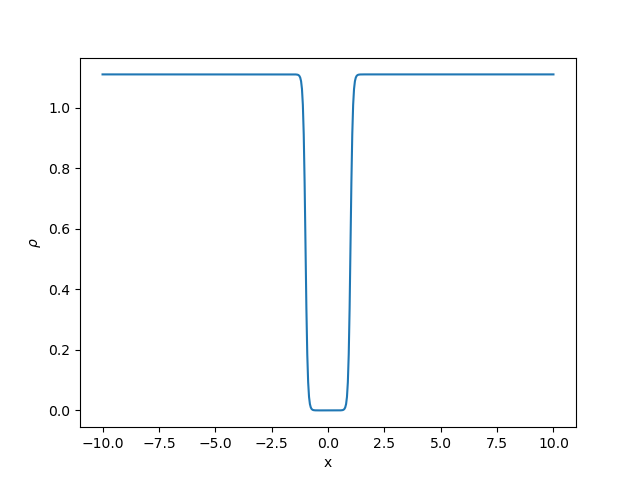}
    \caption{Initial fluid density field for compensated void in one dimension, with an increased density outside the void so that the mean density is 1.}
    \label{fig:rho_IC}
\end{figure}

\begin{figure}

\subfloat[velocity field]{\includegraphics*[width=\linewidth]{1d_void_ic_vel.png} \label{fig:vel_ic}}

\subfloat[velocity potential]{\includegraphics*[width=\linewidth]{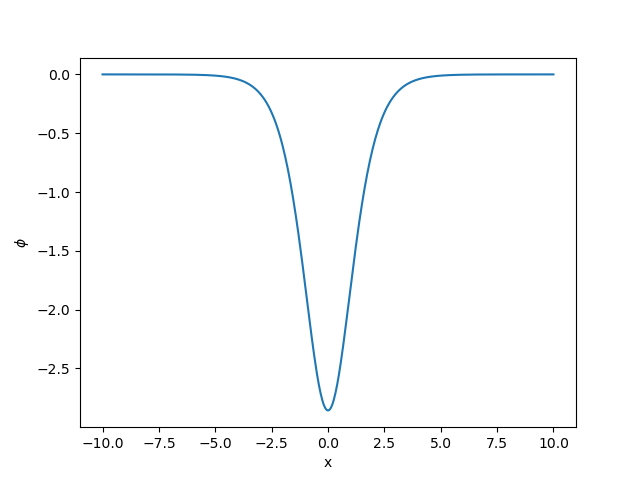}\label{fig:phi_ic}}

\caption{Initial velocity field and its potential.}
\label{fig:vel_and_pot}

\end{figure}

In this Section we will present numerical solutions of the void evolution in one dimension using the SP system. We start by studying the approximate case of a void where the gravitational potential is calculated at each time-step, but not included in the Schr\"odinger equation for the evolution of the void. In other words, the gravitational potential is updated as the fluid moves but the effect of this evolution on the fluid motion is ignored. This has the effect that the fluid moves as a collection of `free particles', a shortcut that has been shown to be remarkably accurate for some applications; see \cite{CS2006a}.
The Schr\"odinger - Poisson system in this approach simplifies to 
\begin{align}
    i \nu \del{\psi}{b} &= -\frac{\nu^{2}}{2} \del{^{2} \psi}{x^{2}}
    \label{eq:1d_schrod}
\end{align}
These equations thus decouple, making the computation much simpler. We begin with a compensated void, shown in Figure \ref{fig:rho_IC}. A compensated void, is such that the mean density of the density field $\rho$ is set to be $1$. To get this shape we use a smoothed $\tanh$ function. 
 We need to set up an initial velocity field in which the fluid particles are moving away from the centre of the void, quickest at the centre and more slowly towards the edges of the void. This behaviour can be seen in Figure \ref{fig:vel_ic}. To get $\phi$, as shown in Figure \ref{fig:phi_ic}, we solve $\nabla \phi = v$. This solution for $\phi$ is only determined up to a constant, which we fix by setting the potential at infinity to be zero (the edge of the box). These initial conditions were chosen to match the phenomenology from the previous solution in Section \ref{analytical_sols}.
 
\begin{figure}

\subfloat{\includegraphics*[width=\linewidth]{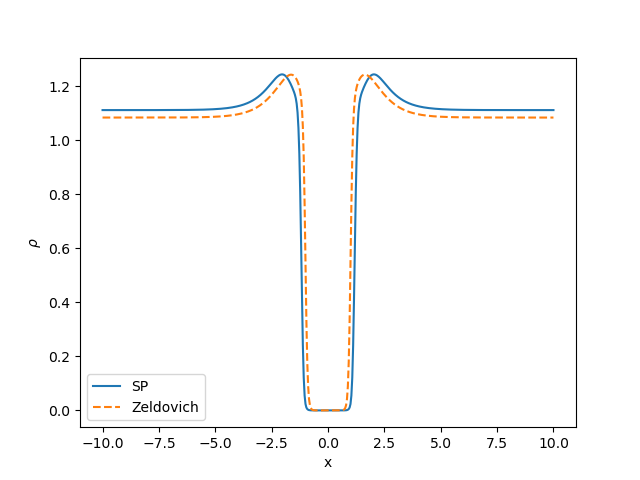} \label{fig:sp_50}}

\subfloat{\includegraphics*[width=\linewidth]{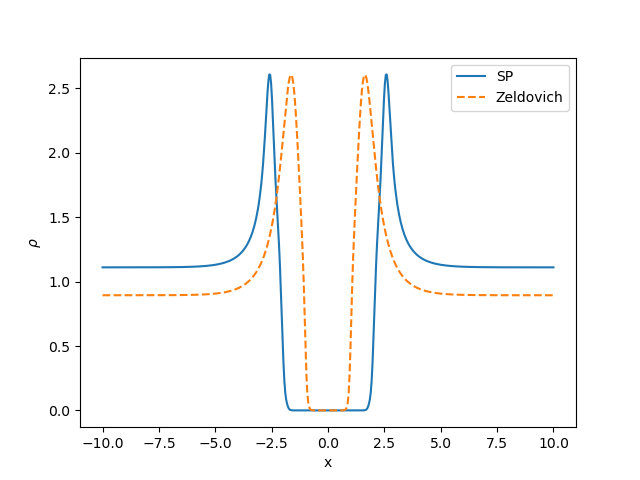} \label{fig:sp_255}}

\subfloat{\includegraphics*[width=\linewidth]{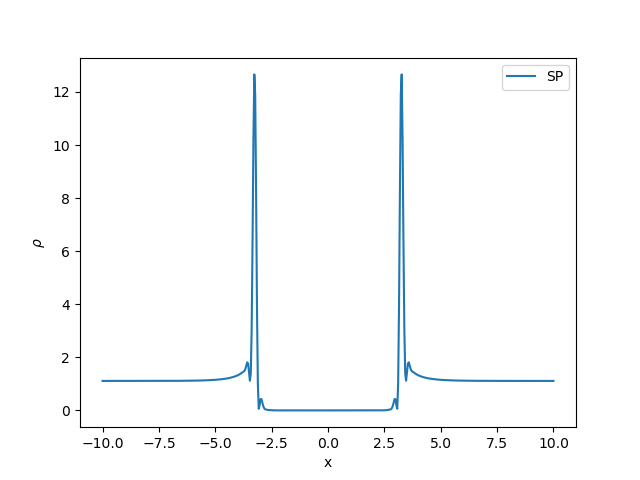} \label{fig:sp_300}}

\caption{Here we are comparing the Schr\"odinger-Poisson formalism for voids with the Zel'dovich approximation applied to the same void, with the Schr\"odinger-Poisson given by the solid blue line and the Zel'dovich approximation given by the orange dashed line. The first plot is an early time-step, where, the second is a late time-step and the final plot is post shell-crossing, where the Zel'dovich approximation breaks down.}

\label{fig:sp_evol}
\end{figure}

 We performed these calculations using periodic boundary conditions. These boundary conditions work quite well with these initial conditions as we have constructed them in such away that the boundary is far enough away from any structure so do not generate any significant cross-boundary interaction as the boundaries are effectively infinitely far away. The Schr\"odinger-Poisson solution is seen in Figure \ref{fig:sp_evol}; this solution is analogous to the Zel'dovich approximation if we work in comoving coordinates $\textbf{x}$, defined by Equation (\ref{eq:comoving}), and time coordinate $b$, as defined by Equation (\ref{eq:btime}), so these two solutions are comparable.

The Zel'dovich approximation is calculated using Equation (\ref{eq:zel_rho_final}), for $\rho_0$ in Figure \ref{fig:rho_IC} and $s(q) = \nabla \phi_0 (q)$, where $\phi_0$ is that in Figure \ref{fig:phi_ic}. Shell-crossing occurs shortly after the second plot in Figure \ref{fig:sp_evol}, where multi-streaming begins and the fluid density field is no longer well defined under Zel'dovich, as it cannot account for multi-streaming. Examining Figure \ref{fig:sp_evol} we see the evolution of the void at three different time-steps in the Schr\"odinger-Poisson and two different time-steps in the Zel'dovich approximation. At early time-step, we see the void begins to expand slightly and a small amount of matter accumulates at the boundary of the void, as seen in Section \ref{void_analytic}. Just before shell-crossing we see that the two models deviate, with Zel'dovich void boundaries not expanding as rapidly, and the SP void is further expanded. In the final time-step, we see that the interference effect caused by multi-streaming is in full effect as the void continues to expand. This interference effect occurs after shell crossing, and is also seen in gravitational collapse described by SP dynamics in, e.g., \cite{Coles_2003} \cite{johnston2009cosmological}  and \cite{Uhlemann_2019}. We think the best way to see what is happening with this model is to watch a movie of this, where the density field changes with time, see ancillary files, which contains this movie. 

\subsection{Void Expansion Speed}

\begin{figure}
    \centering
    \includegraphics[width=\linewidth]{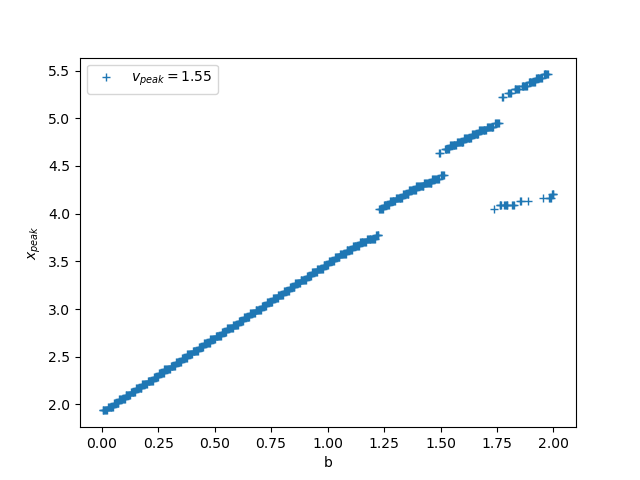}
    \caption{Positions of the peak of the mass at the edge of the void by SP regime as a function of time; the jumps are caused by the formation of subsidiary peaks through wave interference.}
    \label{fig:outward_peak_velocity}
\end{figure}

It can be seen in Figure \ref{fig:sp_evol} that the edges of the void move outwards in comoving coordinates as time goes on. Since the peaks in $\rho$ are sharply defined at least until the strong interference develops after shell-crossing, their positions can be used to define an expansion speed that can be extracted easily from the numerical calculations.  Doing this, we found the group velocity of this peak to be generally constant in time (as defined by $b$), with some jumps once the interference pattern starts to develop. This can be seen in Figure \ref{fig:outward_peak_velocity}, which shows the position of the highest peak at the cusp of the void changes with respect to time. The resulting plot is roughly linear in the coordinates used, which indicates a free expansion of the void region discussed by e.g. \cite{Bert85} and \cite{CB1990}). At very late times one expects a void embedded in a baryonic fluid to enter an adiabatic phase described by the Sedov solution for blast waves (\cite{sedov}), but the physical behaviour of the quantum fluid described here would probably be quite different and in any case we only look at the initial phase of the expansion here.

\subsection{Including the potential term}
\label{with_potential} 

\begin{figure}

\subfloat{\includegraphics*[width=\linewidth]{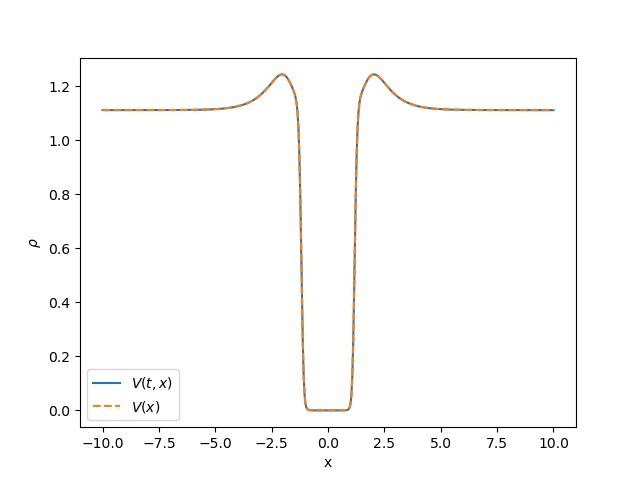} \label{fig:sp_pot_20}}

\subfloat{\includegraphics*[width=\linewidth]{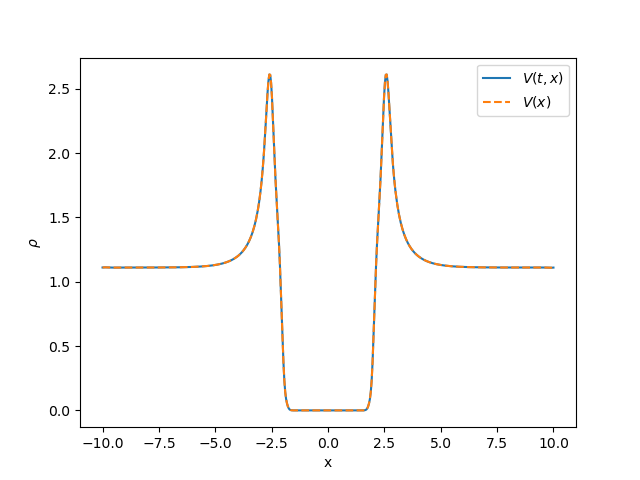} \label{fig:sp_pot_110}}

\subfloat{\includegraphics*[width=\linewidth]{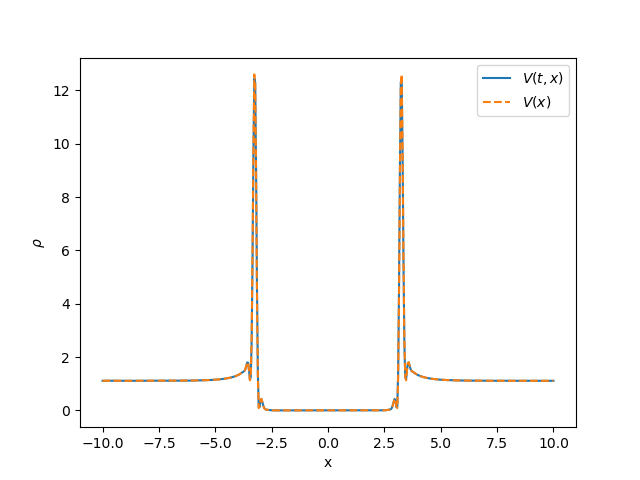} \label{fig:sp_pot_220}}

\caption{Comparing the solution when we include a static potential (dashed orange) and a time dependent potential (solid blue). These plots are the same time-steps as those given in Figure \ref{fig:sp_evol} so that we can compare}
\label{fig:pot_compare}

\end{figure}

In this section we look at including a static potential and a time-varying potential term in the Schr\"odinger-Poisson system of equations. For the static potential, the potential is calculated by Equation (\ref{eq:1d_poiss}) from the initial conditions, and then used as a time-independent constant in 
\begin{align}
    i \nu \del{\psi}{b} &= -\frac{\nu^{2}}{2} \del{^{2} \psi}{x^{2}} + V \psi. 
    \label{eq:1d_schrod_pot}
    \\
    \del{^{2} V}{x^{2}} &= 4\pi G |\psi|^{2}.
    \label{eq:1d_poiss}
\end{align}
For the time-varying potential, both Equations (\ref{eq:1d_schrod_pot}) and (\ref{eq:1d_poiss}) are solved simultaneously as a coupled system of PDEs  for a time-dependent $\psi$ and $V$. Since the evolution of the gravitational potential is quite small, we did not expect either of these extensions to have a drastic effect on the results, but we include them here for completeness. The effect of the potential on the system is only seen at late time-steps and only has an effect on the interference pattern seen after shell crossing. These effects can be seen by comparing Figures \ref{fig:sp_evol} and \ref{fig:pot_compare}. 

The initial conditions show in Figures \ref{fig:rho_IC} and \ref{fig:vel_and_pot} were used for these calculations also, so they can be compared. Comparing Figures \ref{fig:sp_evol} and \ref{fig:pot_compare}, we don't see a large difference between the images, although it is worth noting that for the same time-step in Figure \ref{fig:pot_compare} we see the interference pattern coming into effect sooner than in Figure \ref{fig:sp_evol}. This is expected, as we expect the inclusion of the evolution of the potential to speed up the gravitational effects of the system. It is also worth noting that the interference pattern is almost identical between the static and time-varying potential, \ref{fig:pot_compare}. This shows us that the potential does not change much over this time-scale which in turn validates the use of the static potential or even the free-particle approximation. We do not see a notable difference in the behaviour of the outward peak velocity, when the potential is included. 

\section{Viscosity in a Schr\"odinger fluid}
\label{viscosity} 

It is interesting to remark that the parameter $\nu$ involved in the SP formalism has the same dimensions as a viscosity. This was discussed for example in \cite{CS2006b} in connection with the adhesion model, an extension of the Zel'dovich approximation; see \cite{GSS89}. 

In order to discuss kinematic viscosity in the SP system in more detail one should compare it not with the Euler Equation (\ref{eq:euler}) but with a Navier-Stokes Equation
\begin{equation}
    \del{\textbf{v}}{t} + (\textbf{v}\cdot \nabla ) \textbf{v} + \frac{1}{\rho} \nabla p  -\eta\frac{\nabla^2 \textbf{v}}{\rho} + \nabla \phi = 0 \label{eq:Navier-Stokes},
\end{equation}
in which case the kinematic viscosity $\eta$ arises naturally as a consequence of the quantum potential term discussed above in Section \ref{sp system}, though its behaviour is however somewhat different from that of a classical viscous fluid, specifically because the viscosity varies spatially and can even be negative. 

We will defer a fuller discussion of the origin and role of viscosity in the SP system to future work but, continuing with the simple fluid analogy of the Schr\"odinger-Poisson model, one can suggest defining something analogous to a Reynolds number for this model as
\begin{equation}
    R = \frac{u l}{\nu},
    \label{reynolds_number}
\end{equation}
In normal fluid mechanics (e.g. \cite{fluid_book}), the Reynolds number is a dimensionless quantity formed from $u$ (a velocity), $l$ (a reasonable length-scale for the system) and $\nu$ is the kinematic viscosity of the fluid. We see in Section \ref{sp system} that $\nu$ is essentially a kinematic viscosity. Although $\nu$ is well defined, it is not clear how one would define the other two quantities. There are many velocity quantities present in this model and it is not exactly clear which one we should choose. There is a momentum associated with the wave-function, but this is not uniform in space and it is not clear how one could convert this to a single number. There is also the velocity of the outward peak of the void, $v_{peak}$, a sort of expansion velocity. Although this is not well-defined, it is a single number with dimensions of velocity that captures the behaviour of the void.  There is also the issue of the length-scale to choose for this Reynolds number. It is obvious to choose the size of the void, however, since the void is expanding, this length is not constant. For this we have decided to choose the initial size of the void for our length. This is still somewhat arbitrary as the initial time and therefore size of the void is arbitrarily chosen. 

\begin{figure}

\subfloat[Peak-velocity calculated for fixed $l$ and varying $\nu$.]{\includegraphics*[width=0.85\linewidth]{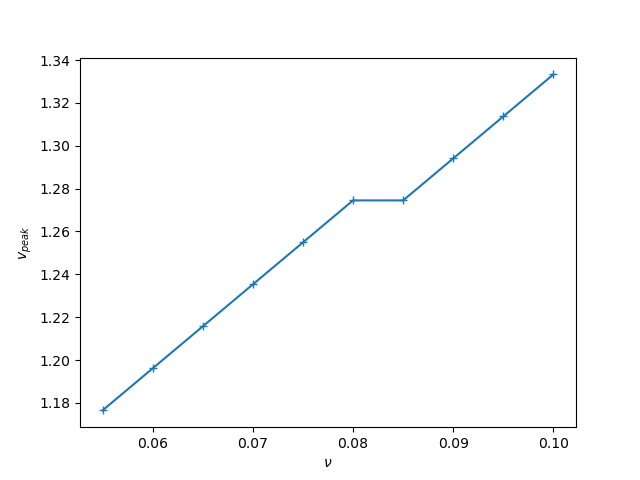} \label{fig:nu_v_corr}}

\subfloat[Peak-velocity calculated for fixed $\nu$ and varying $l$.]{\includegraphics*[width=0.85\linewidth]{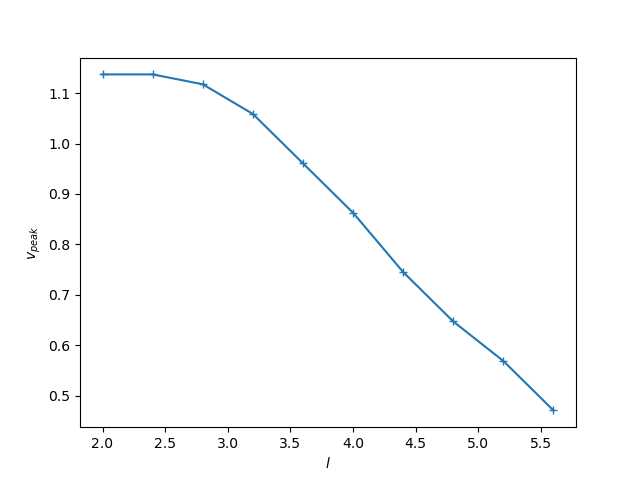} \label{fig:lv_corr}}

\caption{Peak-velocity calculated for (a) fixed $l$ and varying $\nu$ and (b) fixed $\nu$ and varying $l$. } 
\label{fig:corr_nu_l_v}
\end{figure}

To test the reliability of this pseudo-Reynolds number, we looked at the relationships between the quantities used to define it. Since $u$ is the calculated value $v_{peak}$, we started by fixing $l$, and comparing various $\nu$ values with the $v_{peak}$ produced by this. As seen in Figure \ref{fig:nu_v_corr}, it is clear that $\nu$ and $v_{peak}$ are proportional to one another. Then we fixed $\nu$ and calculated $v_{peak}$ for various $l$ values and as seen in Figure \ref{fig:lv_corr}, it is clear that $l$ and $v_{peak}$ have an inversely proportional relationship. Due to the nature of how $v_{peak}$ is calculated it is not possible to fix $u$ and look at the relationship between $\nu$ and $l$, but it suffices for a quick illustration.

The next obvious step is to compute a Reynolds number for the system described above. The initial size of the void can be seen in Figure \ref{fig:rho_IC}. However, it is not very clear in Figure \ref{fig:rho_IC} where the edge of the void is,or what the edge of a void would mean, but we have chosen it, in this case, to be where the density begins to `drop off'. This was defined to be at $\pm 1$, therefore making this length $l = 2.0$. The frequency parameter, or viscosity, was set to be $\nu = 0.05$. As seen above the peak velocity was calculated and found to be $v_{\rm peak} = 1.55$. This would give us a Reynolds number $R = 62.0$. This is equivalent to a void $30~$Mpc wide, with expansion velocity of $27~$ km s$^{-1}$ and a viscosity of $ 4\times 10^{10}~$ km$^{2}s^{-1}$.  The actual value for the effective viscosity is not well known, so we used the value stated and discussed in \cite{viscous} for illustration.

\section{Conclusions and Discussion}
\label{conclusion} 

In this paper we have explored the dynamical evolution of cosmic voids through the lens of the Schr\"odinger-Poisson approach. Following \cite{widrowKaiser1993}, we argue that whether or not dark matter does behave quantum-mechanically, this is an advantageous approximate approach that avoids the formation of caustics in the Zel'dovich approximation outlined in Section \ref{zeldy}. In an era dominated by massive $N$--body simulations there is still space for analytical approaches such as these because they are much less expensive from a computational perspective and allow us to explore parameter space much more quickly.

 Combined with other works using the SP model to study gravitational collapse such as \cite{johnston2009cosmological} and \cite{Uhlemann_2019}, demonstrates that this has much greater potential for the evolution of voids than the the Zel'dovich approximation; see Figure \ref{fig:sp_evol}. The efficacy of this approach is further enhanced by the fact that this model has scaling properties so that computations for voids of different sizes and initial expansion rates can be computed cheaply. 

Obviously isolated voids such as those we have studied here are idealized so their main function is as a testing ground to enable us to develop insight into the behaviour of the SP system.  In future work we plan to explore the evolution of more realistic 3D structures. A particularly interesting potential application of the approximate dynamics deployed here is the problem of cosmic reconstruction, for example from redshift-space to real space, or evolving structure back in time from the present epoch to some earlier state, discussed for example by \cite{reconstruction}.

The above comments apply to our use of the SP dynamics as approximate model for the behaviour of collisionless cold dark matter. If it turns out that dark matter is not cold but instead is for example ultra-light axionic matter then  $\nu$ takes on physical relevance as it is inversely proportional to the mass of the light particle. It is important to stress however that the SP approach has a much wider range of potential use than this specific scenario.

\section*{Acknowledgments}
PC is grateful to William Hunter for discussions during a preliminary investigation of this topic. We thank Cora Uhlemann and Alex Gough for their constructive comments and interesting conversation regarding the Schr\"odinger-Poisson formalsim. We also thank the referees for their very helpful review which aided the presentation of the results in this paper. 

\newpage
\bibliographystyle{mnras}
\bibliography{bibliography}

\end{document}